\documentclass[preprint,authoryear,12pt]{elsarticle}

\usepackage{epsfig}

\usepackage{amssymb}

\usepackage{aas_macros}

\usepackage{color}
\usepackage{array}
\usepackage{amsmath}
\usepackage{soul}
\usepackage{hyperref}

\journal{Astronomy and Computing}

\begin{document}

\begin{frontmatter}



\title{{\sc Smart}: A program to automatically compute  accelerations and
variational equations}


\author[fcag,ialp]{D. D. Carpintero\corref{pie1}}
\author[unrn,conicet]{N. P. Maffione}
\author[iimct,dauls]{F. A. G\'omez}
\cortext[pie1]{Corresponding author: Observatorio Astronómico - Paseo del Bosque s/n, 1900 La Plata, Buenos Aires, Argentina. E-mail address: ddc@fcaglp.unlp.edu.ar}

\address[fcag]{Facultad de Ciencias Astron\'omicas y Geof\'isicas, Universidad
Nacional de La Plata, La Plata, Argentina}
\address[ialp]{Instituto de Astrof\'isica de La Plata, UNLP--Conicet,
La Plata, Argentina}
\address[unrn]{Universidad Nacional de R\'io Negro, R\'io Negro, Argentina}
\address[conicet]{Consejo Nacional de Investigaciones Cient\'ificas y T\'ecnicas, Buenos Aires, Argentina}
\address[iimct] {Instituto de Investigaci\'on Multidisciplinar en Ciencia y Tecnolog\'ia, Universidad de La Serena, Raúl Bitrán 1305, La Serena, Chile}
\address[dauls]{Departamento de Astronom\'ia, Universidad de La Serena, La Serena, Chile}
\begin{abstract}

 Modern astronomical potentials modeling galaxies or stellar systems can be rather involved, and deriving their first derivatives (accelerations) and second derivatives (variational equations) in order to compute orbits and their chaoticity may be a formidable task. We present here a fully automated routine, dubbed {\sc Smart}, with which the accelerations and the variational equations of an arbitrary potential that has been written in the {\sc Fortran 77} language can be computed. Almost any {\sc Fortran 77} statement is admitted in the potential, and the output are standard {\sc Fortran 77} routines ready to use. We validate our algorithm with a set of potentials including time-dependent, velocity-dependent and very complex potentials that even involve auxiliary routines. We also describe with some detail a realistic seven-component Galactic potential, {\sc MilkyWayHydra}, which yields very involved derivatives, thus being a good test bed for {\sc Smart}.
\end{abstract}


\end{frontmatter}


\section{Introduction}
\label{intro}

The chaotic behaviour of an orbit in a given dynamical system can be numerically established through algorithms that fall within two broad categories: those that analise the frequencies of the trajectory \citep[e.g.][]{BS82,L90,SN96,CA98,PL98}, and those based on the evolution of deviation vectors, called variational indicators \citep[e.g.][]{BGS76,BGGS80a,BGGS80b,VC94,CV96,FGL97,VCE99,CS00,SEE00,S01,LF01,FLFF02,CGS03,SESF04,SBA07,MDCG11,DMCG12,DMCG12p,MDCG13,CMD14}. For the sake of completeness, let us also mention the  Poincar\'e surfaces of section \citep[e.g.][]{HH64}, with which a qualitative study of chaos may be done.

To compute any of the variational indicators, the so-called variational equations should be integrated along with the equations of the motion. Whereas the latter includes the first derivatives of the potential with respect to the positions (accelerations), the former need its second derivatives. In cases of potentials that depend on velocities, the second derivatives with respect to them are also needed. As long as the potential is simple, these derivatives are usually tractable, in the sense that they can be readily computed by hand and coded in a program. But modern galactic potentials can be very involved, summing up several complex components, so the task of computing and programming the derivatives can be formidable. In this article, we present an automated algorithm to generate derivatives, the program {\sc Smart}, which can be applied directly on a {\sc Fortran~77} (hereafter F77) coded potential, and that outputs {\sc F77} subroutines for its accelerations and its variational equations.

Despite the fact that the program presented here can be used on its own, our main motivation was to develop this code as part of a larger software package dedicated to the computation of the variational indicators. In that sense, {\sc Smart} was conceived as the second element of the so-called {\sc LP-VIsuite} (La Plata Variational Indicators Suite), an open-source software specifically designed to compute a plethora of chaos indicators based on the evolution of deviation vectors. The suite is composed of three basic elements, namely (a) the {\sc LP-VIcode} (current version 2.0.1; \citealt{CMD14}), which is the kernel code and includes a library with more than ten of the most worldwide used variational indicators, (b) the automatic differentiation pre-processing slave program presented here, {\sc Smart} (current version 1.2.1), and (c) a ready-to-use fully modifiable and realistic seven-component Galactic potential, {\sc MilkyWayHydra} (current version 2.0) that we also use here to validate {\sc Smart}, and was previously used in other studies such as \cite{MGetal18} and \cite{Getal20}. Further details about the suite and its components can be found at:

\begin{center}
\url{http://lp-vicode.fcaglp.unlp.edu.ar/}
\end{center}

The software package is freely available at the website.

\section{Mathematical preliminaries}

Let ${\bf\dot w}={\bf F}({\bf w})$ be the equations of the motion of a test particle in a dynamical system, where ${\bf w}=({\bf x}, {\bf v})$ with {\bf x} the position of the particle, {\bf v} its velocity, and {\bf F} the vector field defining the dynamical system. Let $n$ be the dimension of the configuration space, so that {\bf w} has $2n$ components.

If the motion is driven exclusively by a potential $\Phi({\bf w},t)$, the equations of the motion can be split as
\begin{align}
{\bf\dot x} &={\bf v}, \label{unoa}\\
{\bf\dot v} &=-\nabla_{\bf x}\Phi({\bf w},t), \label{unob}
\end{align}
where the operator $\nabla_{\bf x}$ indicates that the derivatives should be computed only with respect to {\bf x}. In writing a program that computes the trajectories, Eqs. (\ref{unoa}), being independent of the potential, can always be coded directly, whereas for the Eqs. (\ref{unob}) the first derivatives of the potential with respect to the positions are needed. The program {\sc Smart} generates a source code that, given $\Phi$, computes Eq. (\ref{unob}).

On the other hand, whether an orbit is of a regular or chaotic nature can be determined with the aid of the so-called variational indicators. These algorithms seek to determine whether two orbits infinitesimally apart will diverge exponentially or not. To this end, the equations of motion are solved along with the variational equations, that is, the first variations of the equations of the motion:
\begin{equation}
\frac{\rm d (\delta{\bf w})}{\rm d t} = \left.\frac{\partial {\bf F}}
{\partial {\bf w}}\right|_{\bf w} \delta{\bf w},
\label{vareq}
\end{equation}
where $\delta{\bf w}$ is the phase-space deviation between the orbits, and $\partial {\bf F}/\partial {\bf w}$ is the $2n\times 2n$ matrix of derivatives of the components of {\bf F} with respect to the components of {\bf w}. These equations can also be split as
\begin{align}
\frac{\rm d (\delta{\bf x})}{\rm d t} &=\delta\mathbf{v}, \label{dosa}\\
\frac{\rm d (\delta v_j)}{\rm d t} &=\sum_{i=1}^n 
 -\frac{\partial^2\Phi}{\partial x_j \partial w_i} \delta w_i, \quad
 j=1,\dots,n. \label{dosb}
\end{align}
Again, Eqs. (\ref{dosa}) are independent of the potential, so they can be wired into any code, whereas Eqs. (\ref{dosb}) are the ones that {\sc Smart} generates as a routine given the potential $\Phi$.

\section{Description of the code}

The program {\sc Smart} is written in standard F77, except for the common non-standard extensions {\tt DO-ENDDO}, {\tt INCLUDE}, {\tt DOWHILE}, lowercase characters, inline comments, and names longer than 6 characters. All real variables are {\tt DOUBLE PRECISION}.

The program, after reading the name of the file where the potential is coded as a {\tt FUNCTION} with three dummy arguments (time, position and velocity), starts scanning it in order to find all the derivable and no derivable variables.\footnote{In the standard F77 nomenclature, there is a distinction between (scalar) 'variables' and 'arrays'. In this article, we will use 'variable' to refer to both kinds of symbolic names, reserving 'array' for those cases in which a non-scalar variable is specifically meant.} Since the derivatives that are sought are with respect to the positions or the velocities, the derivable variables are those that are the result of some mathematical computation involving positions or velocities, or any of their descendants. If no variable is found that depended on the velocities, {\sc Smart} sets a flag in order to avoid the future computation of derivatives with respect to velocities. 

A second scan is then initiated, in which any statement not involving a derivable variable is copied as is to the output file. If, instead, there are one or more mathematical operations acting on a derivable variable,
each operation is sent in turn to a routine that computes the corresponding code of its derivative. The resulting statement, for each operation, is dumped to the output file. When there are more than one mathematical operation on the same statement, the derivative is computed recursively, using the expression of the derivative of a composition of functions:
\begin{equation}
\frac{\mathrm{d}}{\mathrm{d}x}(f\circ g)(x)\equiv
\frac{\mathrm{d}}{\mathrm{d}x}f(g(x))=
\frac{\mathrm{d}f}{\mathrm{d}g}
\frac{\mathrm{d}g}{\mathrm{d}x}.
\end{equation}

The file with the derivatives is preambled and ended with a set of statements that converts it in a ready-to-use F77 subroutine that computes the negative of the gradient of the potential (i.e., the accelerations).

After the foregoing operations are finished, a new scan computes the corresponding second derivatives, dumping the resulting statements to another file that also is converted to a subroutine that computes the time derivatives of the velocity components of the deviation vector. This pass is done only if the user chooses to do so. For economy of writing, hereinafter we will sometimes call 'pass 1' to refer to the computation of the accelerations, and 'pass 2' to refer to the computation of the derivatives of the deviation vector.

\subsection{Input}

The input parameters, read from the input files {\tt smart.in} and {\tt smart.par}, are only three: (a) the name of the file where the potential is coded (first record of {\tt smart.in}), (b) a digit that indicates whether only the accelerations (digit $=$ {\tt 1}) or also the variational equations (digit $=$ {\tt 2}) are to be computed (second record of {\tt smart.in}), and (c) the dimension $n$ of the potential (first {\tt PARAMETER} of {\tt smart.par}).

\subsubsection{Restrictions on the code of the potential}

The user-provided F77 code to compute the potential should be written in {\tt DOUBLE PRECISION}. Therefore, no {\tt REAL*4} variables are expected to appear. For the same reason, no intrinsic function that converts from and to single precision is expected to be present in the code ({\tt REAL}, {\tt FLOAT}, {\tt SNGL}, {\tt DBLE}, {\tt DPROD}). Complex variables and their corresponding intrinsics {\tt CMPLX}, {\tt AIMAG}, and {\tt CONJG} are not expected in the code either. If the program {\sc Smart} finds any of these functions, it stops with a warning message. 

On the other hand, since the potential should be derivable in order to have a physical meaning, the non-derivable numerical intrinsics are also expected to be absent in the code (truncations: {\tt INT}, {\tt IFIX}, {\tt IDINT}, {\tt AINT}, {\tt DINT}, {\tt ANINT}, {\tt DNINT}, {\tt NINT}, {\tt IDNINT}, transfer of sign: {\tt SIGN}, {\tt ISIGN}, {\tt DSIGN}, positive difference: {\tt DIM}, {\tt IDIM}, {\tt DDIM}, remaindering: {\tt MOD}, {\tt AMOD}, {\tt DMOD}, and maxima/minima ({\tt MAX}, {\tt MAX0}, {\tt AMAX1}, {\tt DMAX1}, {\tt AMAX0}, {\tt MAX1}, {\tt MIN0}, {\tt AMIN1}, {\tt DMIN1}, {\tt AMIN0}, {\tt MIN1}). If the program {\sc Smart} finds any of these functions, it stops with a message.

The code of the potential routine should begin thus (aside from comment lines or lowercase/uppercase spelling):
\begin{verbatim}
       FUNCTION pot(t,x,n)
       INTEGER n
       DOUBLE PRECISION pot,t,x(n)
\end{verbatim}
That is, the routine should be a {\tt FUNCTION} and its name should be {\tt pot}. The three dummy arguments should be named and given in the order as shown. They are: the time coordinate {\tt t}, that must be present even if the potential is time independent; an array {\tt x} containing the Cartesian coordinates ${\bf x},{\bf\dot x}$ (in that order) of the point of the phase space  at which the potential is to be evaluated, and the dimension of the phase space {\tt n}. Note that if the potential is independent of the velocities, the passing-by-address feature of F77 allows sending only the first half of {\tt x}. The {\tt DOUBLE PRECISION} declaration may be replaced by the non-standard {\tt REAL*8} if the compiler accepts it. Any additional parameter should be passed to the routine by {\tt COMMON}.

After the foregoing statements, the rest of the routine may be any valid F77 code, including slave subprograms and the  non-standard extensions {\tt DO-ENDDO}, {\tt INCLUDE}, {\tt DOWHILE}, lowercase characters, inline comments, and names of variables longer than 6 characters, except for the following seven limitations:

\begin{enumerate}
\item All the variables, with their dimensions if any, should be declared with a specification statement specifying their type, as if {\tt IMPLICIT NONE} were in force. Therefore, neither the {\tt DIMENSION} nor the {\tt IMPLICIT} statements should be used. Any number of variables can be specified in a given specification statement.

\item The names of variables, including the declaration of their dimensions if any, should be less than 40 characters long.

\item The names of variables must not begin with the letter {\tt o} (lowercase or uppercase). This initial letter is reserved for variables generated by {\sc Smart}.

\item Arithmetic and logical {\tt IF}s are not allowed. They should be replaced with block {\tt IF}s, i.e., the keyword {\tt THEN} should always be present.

\item Statement function statements are not allowed. They should be coded as normal statements in the body of the routine.

\item The dummy array {\tt x} should not appear with a variable index (e.g. {\tt x(k)}), but always with an explicit number (e.g. {\tt x(1)}). This is by far the most restrictive limitation, but it is always achievable with a little bit of additional code.

\item No auxiliary (slave) subprogram should use derivable variables.
\end{enumerate}

The {\sc Smart} program detects whether any of these restrictions are violated, and in that event it stops with a warning message.

\subsection{Output}

The output consists of a file containing an F77 subprogram for the computation of the accelerations corresponding to the input potential, and an optional second file containing an F77 subprogram for the computation of the respective variational equations.

The initial statements of the first subprogram are:
\begin{verbatim}
**************************************************************
      SUBROUTINE acelera(t,x,n,acc)
      DOUBLE PRECISION acc(n/2)
      INTEGER n
      DOUBLE PRECISION pot,t,x(n)
**************************************************************
\end{verbatim}

The meaning of the {\tt t}, {\tt x}, and {\tt n} input dummy arguments is the same as in the potential subprogram. The {\tt acc} output dummy argument returns to the calling program the components of the acceleration vector.

The first statements of the second subprogram are:
\begin{verbatim}
**************************************************************
      SUBROUTINE variac(t,x,dx,n,dax)
      DOUBLE PRECISION dx(n),dax(n/2),var(n/2,n)
      DOUBLE PRECISION acc(6/2)
      INTEGER n
      DOUBLE PRECISION pot,t,x(n)
**************************************************************
\end{verbatim}

Here {\tt t}, {\tt x}, and {\tt n} are again input dummy arguments with the same meaning as in the potential subprogram. The additional input array {\tt dx} contains the components of the deviation vector. The output array {\tt dax} returns the derivatives with respect to the time of the deviations of the velocity. Note that the array {\tt acc} is not a dummy argument, and therefore the standard F77 forbids the use of {\tt n} in defining its dimension. That is why the number {\tt 6} appears in this example, which was taken from the output corresponding to a potential with a 6D phase space. Other (even) dimensions can appear according to the input potential. 

Both output routines are complete F77 subprograms, ready for use.

\subsection{Details of the algorithm of derivation}

The lexical analysis of each F77 statement is always done taking into account that blank spaces can be freely interspersed, even inside a keyword, and that lowercase and uppercase letters are equivalent. Hereinafter, a 'real variable' is understood to mean a 'double precision real variable'.

\subsubsection{Derivable variables}

The first task of {\sc Smart} is to load a list of variables used in the computation of the potential, with their respective dimensions if any. Since all variables should be declared, it suffices to scan the code only until the first executable statement is reached. This also allows to classify from the outset whether a variable is real, integer, logical or character; only the first ones are of interest, for only them can be derivable variables. After this first step, the rest of the code is scanned to determine whether each real variable is derivable or not. In this respect, let us emphasize that to compute the derivative of an F77 {\tt FUNCTION}, the only statements that have to be processed are the assignments to a real variable. Any other statement, be it a declaration, a structural statement (e.g. {\tt DO}, {\tt IF}, etc.), a logical or integer assignment, etc., can be ignored for derivation purposes. Therefore, each statement is read in turn, with its continuation lines if any, to determine whether it corresponds to a real assignment. To this end, it is first scanned to detect the presence of an equal sign. If it is present, the statement is further analized to detect if it is any of the non-assignment F77 statements that may contain that sign: {\tt  OPEN}, {\tt CLOSE}, {\tt INQUIRE}, {\tt READ}, {\tt WRITE}, {\tt PRINT}, {\tt  BACKSPACE}, {\tt ENDFILE}, {\tt  REWIND}, {\tt PARAMETER}, {\tt DATA}, and {\tt DO}. If the statement is not any of the above, it is additionally scrutinized to verify whether its left hand side is a real variable. Any statement that satisfies all these conditions is passed to the next algorithm; otherwise, it is discarded and the next one is read.

Once a real assignment statement is detected, its real variables are classified according to the following criteria:
\begin{itemize}
\item {\tt pot} and {\tt x} are stored as derivable from the outset; {\tt t} is stored as non-derivable.
\item If the first appearance of a new variable is in the right hand side (rhs) of an assignment, it only can be a non-derivable variable, for its value should have already be set (that is, it can only be a variable that has come through a {\tt COMMON}).
\item If the first appearance of a new variable is in the left hand side (lhs) of an assignment, it is derivable if there are derivable variables in the rhs; otherwise, it is non-derivable.
\item Subsequent appearances of any variable in a rhs or a derivable variable in a lhs do not change their condition.
\item Subsequent appearances of a non-derivable variable in a lhs make it change to derivable if there are derivable variables in the corresponding rhs.
\end{itemize}

\subsubsection{Lexical analysis of the statements}

Once the derivable variables were identified, the potential routine is scanned again, from the first executable statement, in order to dissect each real assignment into its atomic constituents (Table \ref{atomic}).

\begin{table}
 \caption{Atomic elements that can appear on a real assignment in F77.}
 \label{atomic}
 \begin{center}
 \begin{tabular}{l>{\tt}l}
  \hline
  Type & {\rm element}\\
  \hline
  algebraic &+ - * / ** = \\
  intrinsic &ABS SQRT EXP LOG LOG10 SIN COS TAN ASIN \\
  & ACOS ATAN SINH COSH TANH ATAN2\\
  constant &{\rm integer or real numbers}\\
  name & {\rm names of integer or real constants ({\tt PARAMETER}s)} \\
  & {\rm names of integer or real variables}\\
  other & , ( ) ! {\rm newline} {\rm label}\\
  \hline
 \end{tabular}
 \end{center}
\end{table}

Each atom is loaded on a stack (array) right away, except for: (a) a minus sign, which is previously classified as binary or unary, and transformed to a $-1$ factor in the latter case if it is not preceding a number, (b) a specific name of an intrinsic function (i.e., an intrinsic with an {\tt A}, {\tt D} or {\tt C} prefixed to its generic name), from which the first letter is removed to keep only the generic name,\footnote{A widespread belief asserts that {\sc F77} intrinsic functions operating on double precision objects should be used with their specific names beginning with {\tt D}. It is not true.} and (c) a statement label, which is immediately output as a {\tt CONTINUE} statement with the same label and is no further processed. A newline or a {\tt !} character (i.e. the start of an inline comment) indicates the end of the statement.

The loading of atoms is temporarily stopped if one of these conditions is met: (a) the loaded atom is an entire operation ({\tt +} or binary {\tt -}) and the previous operation loaded in the stack is an algebraic operation ({\tt +}, binary {\tt -}, {\tt *}, {\tt /} or {\tt **})\footnote{By a slight abuse of language, we call the power an algebraic operation even when the exponent is not an integer.}; (b) the loaded atom is a rational operation ({\tt *} or {\tt /}) and the previous operation loaded in the stack is a power or a rational operation; (c) the loaded atom is a right bracket and the previous operation loaded in the stack is an algebraic operation or an intrinsic. In all these cases a mathematical operation is being closed, and the derivation process is applied to it before a new atom is loaded. Once the derivation process is ended, the operation and its operands are removed from the stack and replaced by their result, the loading of atoms is resumed, and the foregoing algorithm is repeated until there are no more atoms to read in the statement. At this point, any pending mathematical operation is resolved (i.e., derived), and the next statement is read. The whole process is repeated until the {\tt END} statement is reached, and then {\sc Smart} simply copies the remaining lines (i.e. slave routines), if any, into the output. Notice that if the routine of the potential and that of the accelerations and/or variational equations are to be present in the same program, these duplicates of the slave routines should be deleted.

\subsubsection{Derivation procedure}

Upon reception of a mathematical operation and its operands by the derivation routine, it first reconstructs the operation in the form of an F77 statement, since its result is usually needed in the derivative. If the operation is the last one of a statement of the potential routine, the lhs of the statement of this partial result is the original variable into which the result is stored. Otherwise, the lhs is defined as the array element {\tt o(}$x${\tt )} (pass 1) or {\tt oq(}$x${\tt )} (pass 2), where $x$ are consecutive numbers; this array element is of course considered a derivable variable from there on. In either case, the statement is output and the atomic operation and its operands are then subjected to the derivative process. 

For both, the two operands and the left member, a name of an array which will contain their derivatives with respect to positions (and possibly to velocitites) is generated. Table \ref{deriv} summarizes the different cases. Note that the entry 'any other' in the Table cannot happen in the case of the left member. Note also that the derivative of an {\tt o(}$x${\tt )} or an {\tt oq(}$x${\tt )} variable will be prefixed with an additional {\tt o} or an additional {\tt oq}, respectively. When performing this step, the argument of an intrinsic function is considered as the left operand, and a right operand with zero derivative is assumed, except in the case of {\tt ATAN2}, which takes two operands. Then, the mathematical derivation is performed on the operation itself. This is explicitly coded for each kind of operation and intrinsic by using the rules of derivation in each case.

\begin{table}
 \caption{Results and arrays for the derivatives. The {\tt //} operation symbolizes the concatenation of strings.}
 \label{deriv}
 \begin{center}
 \begin{tabular}{lll}
  \hline
  {\rm the derivative of...} & {\rm generates...}\\
  \hline
  \tt x   & {\tt 1} or {\tt 0} \\
  \tt pot & \tt acc\\
  \tt acc & \tt var\\
  name of derivable variable (pass 1)& {\tt o//}{\it name}\\
  name of derivable variable (pass 2)& {\tt oq//}{\it name}\\
  any other & {\tt 0}\\
  \hline
 \end{tabular}
 \end{center}
\end{table}

A simple example will help to understand the procedure. For the sake of brevity, let us suppose that the potential depends only on the Cartesian coordinates $(x,y)$. If the following statements are to be processed:
\begin{verbatim}
      a = b*x(1) + x(2)
      c = SIN(a)
\end{verbatim}
where {\tt b} is a non-derivable variable, then the first atomic operation will produce the following code:
\begin{verbatim}
      o(1) = b*x(1)
\end{verbatim}
and its derivatives, using the rules of derivation and the results of Table \ref{deriv}, would be coded
\begin{verbatim}
      oo(1,1) = b*1 + 0*x(1)
      oo(1,2) = b*0 + 0*x(1)
\end{verbatim}
Then, the second atomic operation of the first statement will produce
\begin{verbatim}
      a = o(1) + x(2)
\end{verbatim}
with derivatives
\begin{verbatim}
      oa(1) = oo(1,1) + 0
      oa(2) = oo(1,2) + 1
\end{verbatim}
Since the second statement is an atomic operation in itself, it will not generate any new {\tt o($x$)} variable, and it will be copied as is. Its derivatives will be:
\begin{verbatim}
      oc(1) = COS(a)*oa(1)
      oc(2) = COS(a)*oa(2)
\end{verbatim}

As already said, once each elemental derivative is completed the stack is updated by replacing the entire operation by its result. The derivation process continues until the {\tt END} statement of the potential file is reached.

To reduce useless code, each statement is simplified before its output in the following cases: (a) any term containing a zero factor is deleted, (b) any unit factor in a product is deleted, (c) any unit exponent is eliminated, leaving only the base.

\subsubsection{Final statements}

In pass 1, the routine generated by the foregoing algorithm is ended with the lines
\begin{verbatim}
      DO oi=1,n/2
         acc(oi)=-acc(oi)
      ENDDO
\end{verbatim}
in order to return to the calling program the negative of the gradient of the potential, i.e., the accelerations. For pass 2, the routine is ended with the lines
\begin{verbatim}
      DO oi=1,n/2
         dax(oi)=0d0
         DO oj=1,n
            dax(oi)=dax(oi)+var(oi,oj)*dx(oj)
         ENDDO
      ENDDO
\end{verbatim}
which compute the temporal derivative of the velocity components of the deviation vector from the second derivatives of the potential. If the potential does not depend on the velocities, the inner loop goes only up to {\tt n/2}.

\section{Validation of the code}

\subsection{Validation of the code using known potentials}

\begin{figure}
\includegraphics[width=0.90\textwidth]{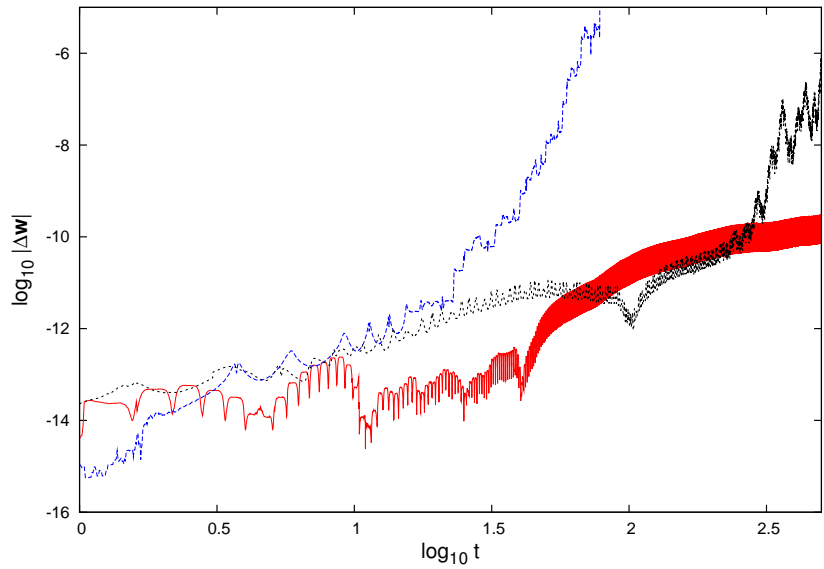}
\caption{Distance $|\Delta \bf{w}|$ in the phase space, as a function of time $t$, between orbits computed with manually coded accelerations and with accelerations obtained with {\sc Smart}. The orbits belong to the Binney potential with $q=0.9, R_{\rm c}=0.14, v_0^2=1, R_e=3$, with initial conditions $(x,y,\dot x,\dot y)=(0.1,0,0.5,0.02)$ (regular orbit, red), $(x,y,\dot x,\dot y)=(0.1,0.5,0,1)$ (chaotic orbit, blue), and $(x,y,\dot x,\dot y)=(0.077,0,0,1.73203)$ (sticky orbit, black).}
\label{orbits}
\end{figure}

The acceleration part of the code was validated with an assorted set of potentials (Table \ref{potes}), ranging from the most elemental (harmonic, Keplerian) to the most involved (Hernquist \& Ostriker, perfect ellipsoid, see references on Table \ref{potes}). The latter require extra routines and even numerical integrations of functions to compute the potential. A time dependent potential (logarithmic with a variable flattening) and a velocity dependent potential (global bifurcation) were also included. For each potential, various initial conditions were chosen with which orbits were computed with the {\sc LP-VIcode}, both with the accelerations computed and coded by hand, and with the accelerations obtained with {\sc Smart}. In all regular cases the orbits thus obtained were the same, aside from the expected numerical differences when the same algorithm is coded in two different ways.  Chaotic orbits, on the other hand, remained the same only until the exponential divergence makes the orbits differ. This is, again, the expected behaviour when the same trajectory in the chaotic region of a potential is integrated with different codes. We also probed sticky orbits, i.e. orbits which behave as regular for a certain time, but then they show themselves as chaotic. As expected, the differences between the manual coding and {\sc Smart} were similar to the regular case when the orbits behaved regularly, and to the chaotic case thereafter. Fig. \ref{orbits} shows the distance in phase space between orbits of the Binney potential computed with accelerations coded by hand and by {\sc Smart}, for a regular, a chaotic, and a sticky case.

\begin{table}
 \caption{Potentials with which the acceleration was validated. Symbols in the expressions, aside from the coordinates $x$, $y$, and $z$, are parameters of the corresponding potential.}
 \label{potes}
 \begin{center}
 \begin{tabular}{lll}
  \hline
  Name & Dim & Reference or expression \\
  \hline
  Binney & 2D & \cite{B82} \\
  Global bifurcation & 1D & \cite{J89}, p. 300 \\
  Harmonic & 1D & $\frac{1}{2}\omega^2 x^2$ \\
  & 3D &  $\frac{1}{2}(\omega_x^2 x^2 + \omega_y^2 y^2 + \omega_z^2 z^2)$\\
  Quartic & 1D & $\frac{1}{2} x^2 + \frac{1}{3} x^3 + \frac{1}{4} \epsilon x^4$ \\
  H\'enon-Heiles & 2D & \cite{HH64} \\
  Hernquist & 3D & \cite{H90} \\
  Hernquist-Ostriker & 3D & \cite{HO92} \\
  Keplerian & 3D & $-GM/\sqrt{x^2+y^2+z^2}$ \\
  Logarithmic & 2D & $\frac{1}{2}v_0^2 \ln(x^2+y^2/q^2+R_{\rm c}^2)$ \\
  Logarithmic, time dependent & 2D & same as above, with \\
  && $q(t)=A+B\exp(-(t-C)^2/2)$ \\
  Logarithmic quadrupolar & 3D & \cite{CW06} \\
  Merritt-Fridman & 3D & \cite{MF96} \\
  Miyamoto-Nagai (MN) & 3D & \cite{MN75}\\
  MN + Hernquist + NFW & 3D & \cite{GHBL10} \\
  MN + Plummer & 3D & see individual entries \\
  MN + Plummer, merid. plane & 2D & \cite{ZC13} \\
  NFW, triaxial & 3D & \cite{VWHS08} \\
  Perfect ellipsoid & 3D & \cite{dZ85} \\
  Plummer & 3D & \cite{P11} \\
  Rational quadrupolar & 3D & \cite{MCW05} \\
  Satoh, prolate & 3D & \cite{CMW99} \\
  Schwarzschild & 3D & \cite{S93} \\
 \hline
 \end{tabular}
 \end{center}
\end{table}

No rotating potential was considered in the validation. This is due to the fact that, in non-inertial frames, there are non-inertial ('fictitious') forces that are not derived from the gradient of a potential. Whereas any centrifugal force can be masked as the gradient of a centrifugal potential, Coriolis forces cannot, and thus are not suitable for {\sc Smart}. However, in these cases the user could add the corresponding accelerations by hand.

\begin{figure}
\includegraphics[width=0.49\textwidth]{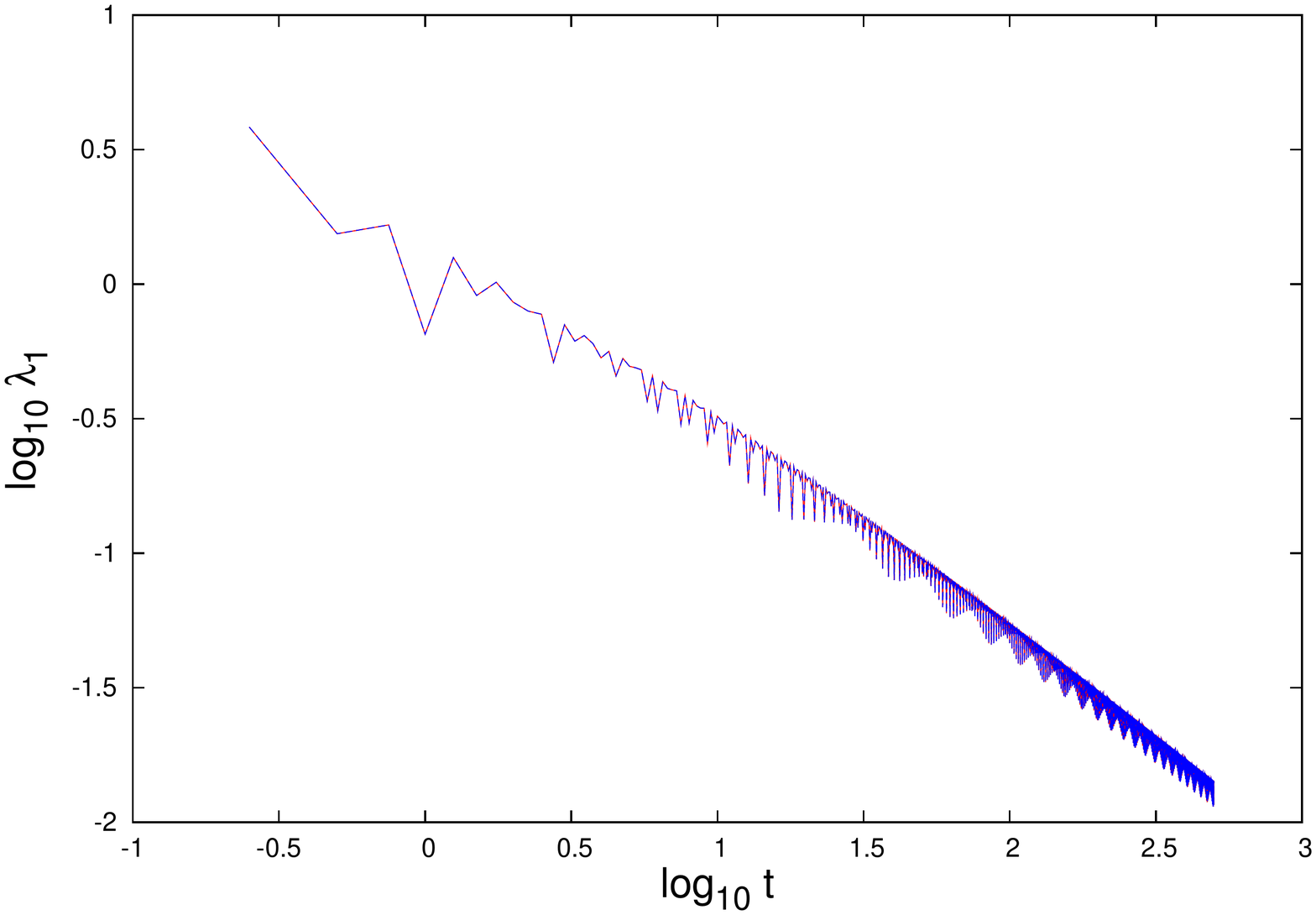}
\includegraphics[width=0.49\textwidth]{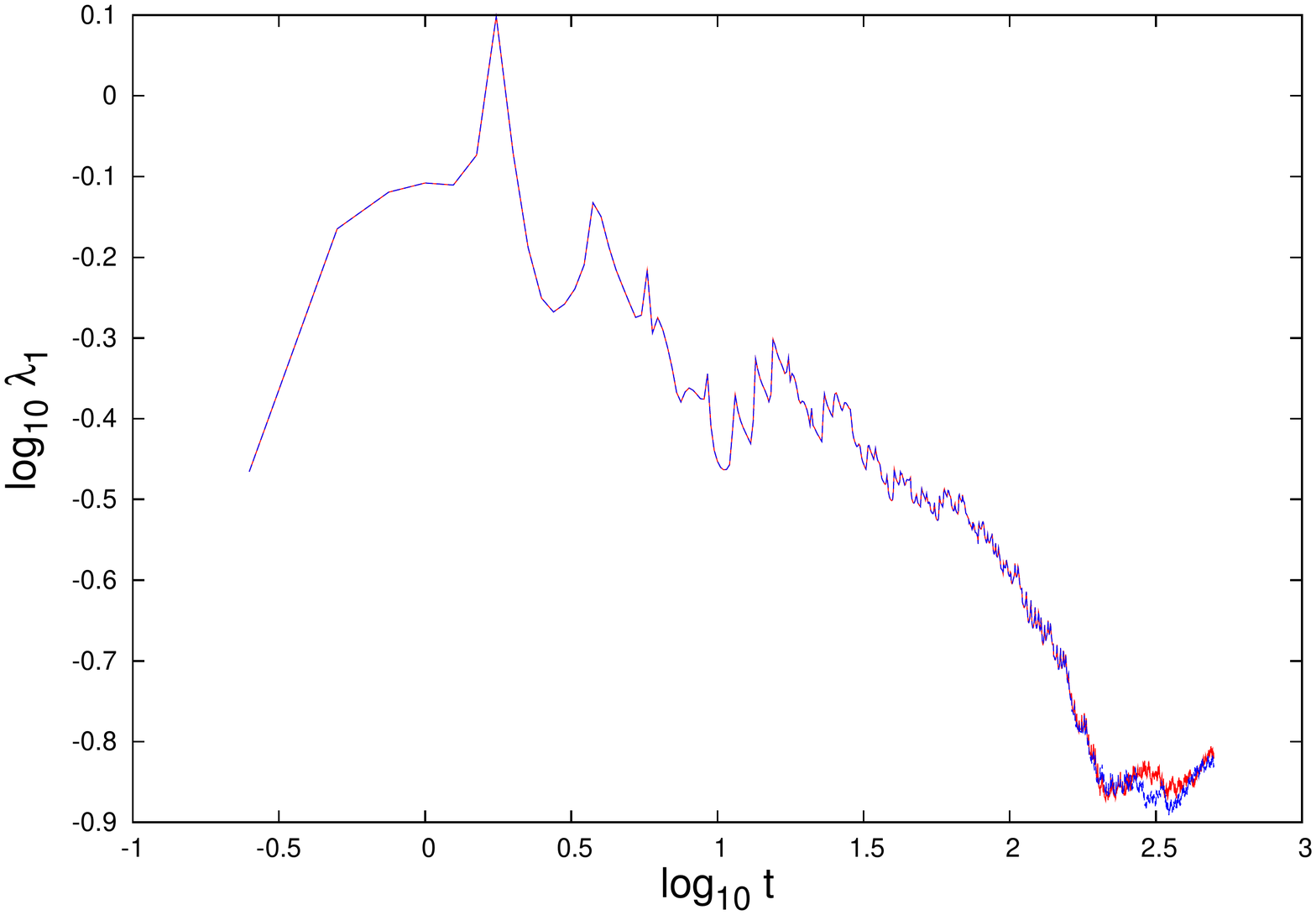}
\caption{Left: The maximal Lyapunov exponent $\lambda_1$ (a chaos indicator) of the regular orbit of Fig. \ref{orbits} (left). The indicator $\lambda_1$ should tend to 0 with time for a regular orbit; in a log-log plot it should be a straight line with negative slope. The blue and red curves correspond to manually coded variational equations and to these equations obtained with {\sc Smart}, respectively. Right: The same but for the chaotic orbit of Fig. \ref{orbits} (right). The curves split when the chaoticity begins to show. The chaotic nature of the orbit is nevertheless found in both cases.}
\label{lyap}
\end{figure}

The variational equations, on the other hand, were validated with a more restricted set of potentials, due to the difficulty in writing down these equations for most potentials (which is, incidentaly, the main reason for developing {\sc Smart}). The chosen potentials were: global bifurcation, H\'enon-Heiles, logarithmic (with a null core radius), and NFW triaxial, all of which admit chaotic regions. We repeated the procedure described for the accelerations, plus the variational equations coded by hand and coded by {\sc Smart}. With them, we computed several chaos indicators with {\sc LP-VIcode} for each orbit. The results were similar to those obtained in the case of accelerations: no differences were found between codes for the regular orbits; the chaotic ones showed slight differences, especially from the moment the indicators showed their chaotic condition, and sticky orbits showed both regimes. In any case, the regular or chaotic quality was always preserved. Fig. \ref{lyap} shows examples of a regular and a chaotic case.

In addition to the above, we run a number of experiments to test the computing times of the manually coded accelerations and variational equations, against those generated with {\sc Smart}, for several of the potentials used to validate the code. We used an Intel(R) Core(TM) i7-4790K CPU (4 cores, 8 threads) @ 4.00 GHz with 8 GB RAM, Windows 7 Ultimate Service Pack 1, 64 bits, running a virtual machine Oracle VM VirtualBox using Ubuntu 18.04 64 bits with 4 GB RAM. The codes were compiled with {\tt gfortran}, with the optimization flag {\tt O3} on. 

Table \ref{tiempos} shows the outcome. The reported times correspond to the integration of $50\,000$ orbits on each potential until an arbitrary final time of 6.135 time units. For the accelerations, the manual coding is, on average, 15 per cent faster than that produced by the {\sc Smart} coding, though individual entries range from 2 per cent to 46 per cent. However, there is no clear correlation between the complexity of the potential and the additional time took by {\sc Smart}.

On the other hand, with the variational equations the manual coding is between 35 per cent and 60 per cent faster than {\sc Smart}, except for the H\'enon-Heiles case in which both codes took almost the same time.

In view of these figures, when orbits need to be computed on big samples of initial conditions, the handcrafted versions of the accelerations and variational equations are worth using over the automatic differentiation version provided by {\sc Smart}. In all other cases, {\sc Smart} is the right choice due to the almost nil effort from the user to set things up.

\begin{table}
 \caption{Comparison of integration times between manual and {\sc Smart} codings. EQM stands for equations of motion (accelerations); VEQ stands for variational equations (plus EQM). Times are given in minutes:seconds.}
 \label{tiempos}
 \begin{center}
 \begin{tabular}{lrrrrrr}
  \hline
  Potential & manual & {\sc Smart} & & manual & {\sc Smart} & \\
   & EQM & EQM & \% & VEQ & VEQ & \% \\ 
  \hline
  Binney & 2:35 & 3:19 & 22.1\\
  Harmonic 3D & 2:40 & 2:55 & 8.8\\
  Hénon-Heiles & 2:03 & 2:07 & 2.9 & 14:05 & 14:16 & 1.2\\
  Hernquist & 2:45 & 3:07 & 12.1\\
  Keplerian 3D & 3:18 & 4:00 & 17.3\\
  Logarithmic & 4:20 & 6:18 & 31.1 & 52:30 & 81:17 & 35.4\\
   Logarithmic, \\
  time dependent & 2:18 & 2:45 & 16.3\\
  Logarithmic quad. & 7:07 & 7:16 & 2.0 \\
  Merritt-Fridman & 5:44 & 9:51 & 45.8 \\
   MN + Hernquist \\
  + NFW & 6:20 & 6:28 & 2.1\\
   NFW triaxial & 9:11 & 9:59 & 12.5 & 394:31 & 630:00 & 38.4\\
   Perfect ellipsoid & 13:38 & 17:55 & 23.9 & 243:03 & 582:22 & 58.3 \\
  Schwarzschild & 2:46 & 2:50 & 2.6\\
 \hline
\end{tabular}
 \end{center}
\end{table}

\subsection{Validation of the code using {\sc MilkyWayHydra}}
\label{S4.1}
{\sc MilkyWayHydra} ({\sc MWH} for short) is the third element of the {\sc LP-VIsuite}: a ready-to-use fully modifiable and realistic multi-component 3D galactic potential. The {\sc MWH} has been conceived mainly to represent late-type galaxies and, in particular, Milky Way-type galaxies. Nevertheless, it can be used to represent early-type galaxies as well, due to its straightforward programming style. The potential is written in standard  {\sc F77} following the rules specified by the suite's latest version of the kernel code ---the  {\sc LP-VIcode}---, and each of the galactic components are clearly identified by using separated blocks. The result is a set of subprograms in a file called {\tt milkywayhydra.pav}, which includes the computation of the potential, the accelerations and the first variational equations (manually coded and checked with symbolic manipulators) ready to use with (exclusively) the {\sc LP-VIcode}. Current version {\sc MWH} 2.0 describes a galactic potential with the following components: a nuclear region, a peanut-shaped bulge, spiral arms, a disk, and a dark matter halo.

In this last validation experiment we test {\sc Smart} with the most demanding (by far) potential we have at hand. Therefore, we are using the default {\sc MWH} 2.0 configuration, the so-called {\sc Hydra} 2.0 which is a realistic Milky Way-type galactic potential meant to provide an accurate representation of the Milky Way potential (for further details on the potential see~\ref{A1}). In particular for the present set of tests, we are using two flavours of the {\sc Hydra} 2.0, i.e. the time independent and the full time dependent versions.

\begin{figure}
\includegraphics[width=0.49\textwidth]{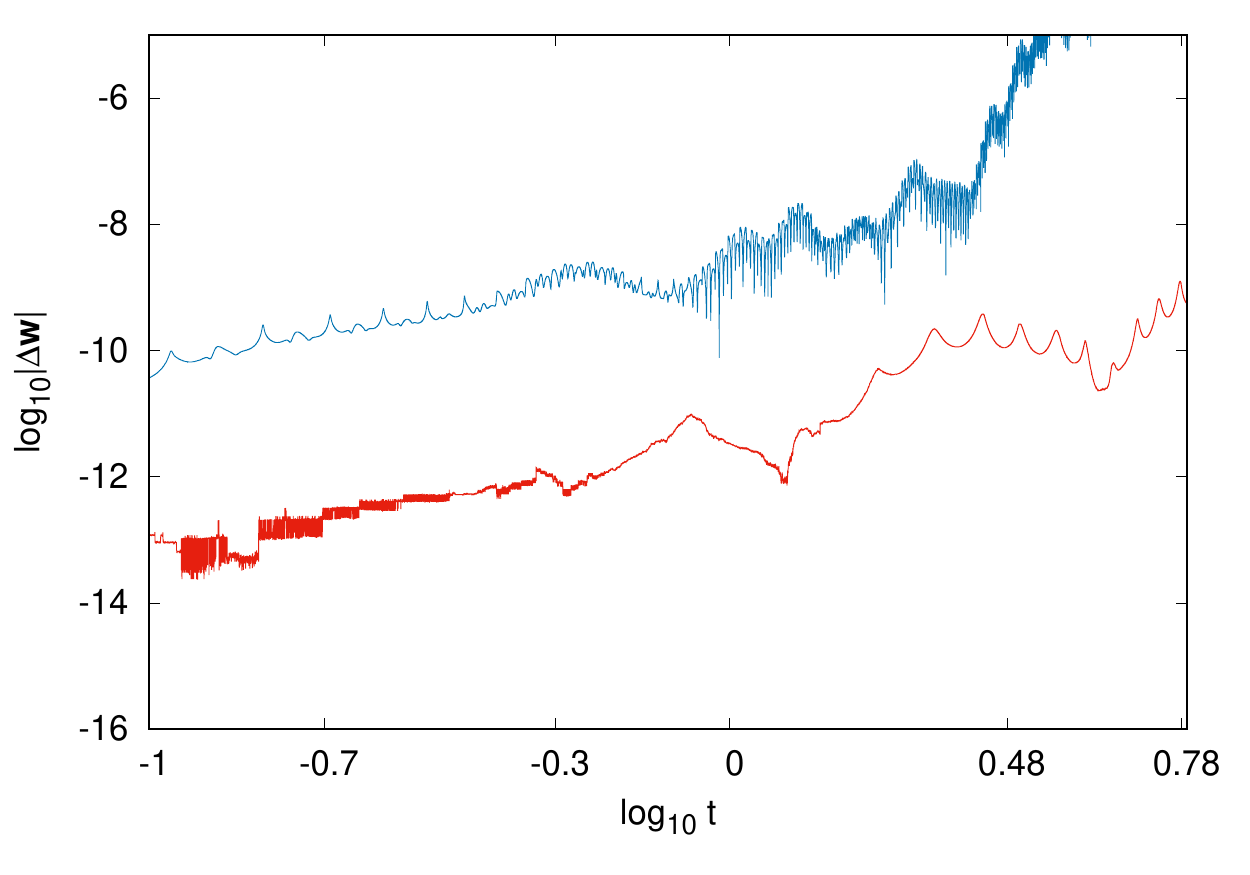}
\includegraphics[width=0.49\textwidth]{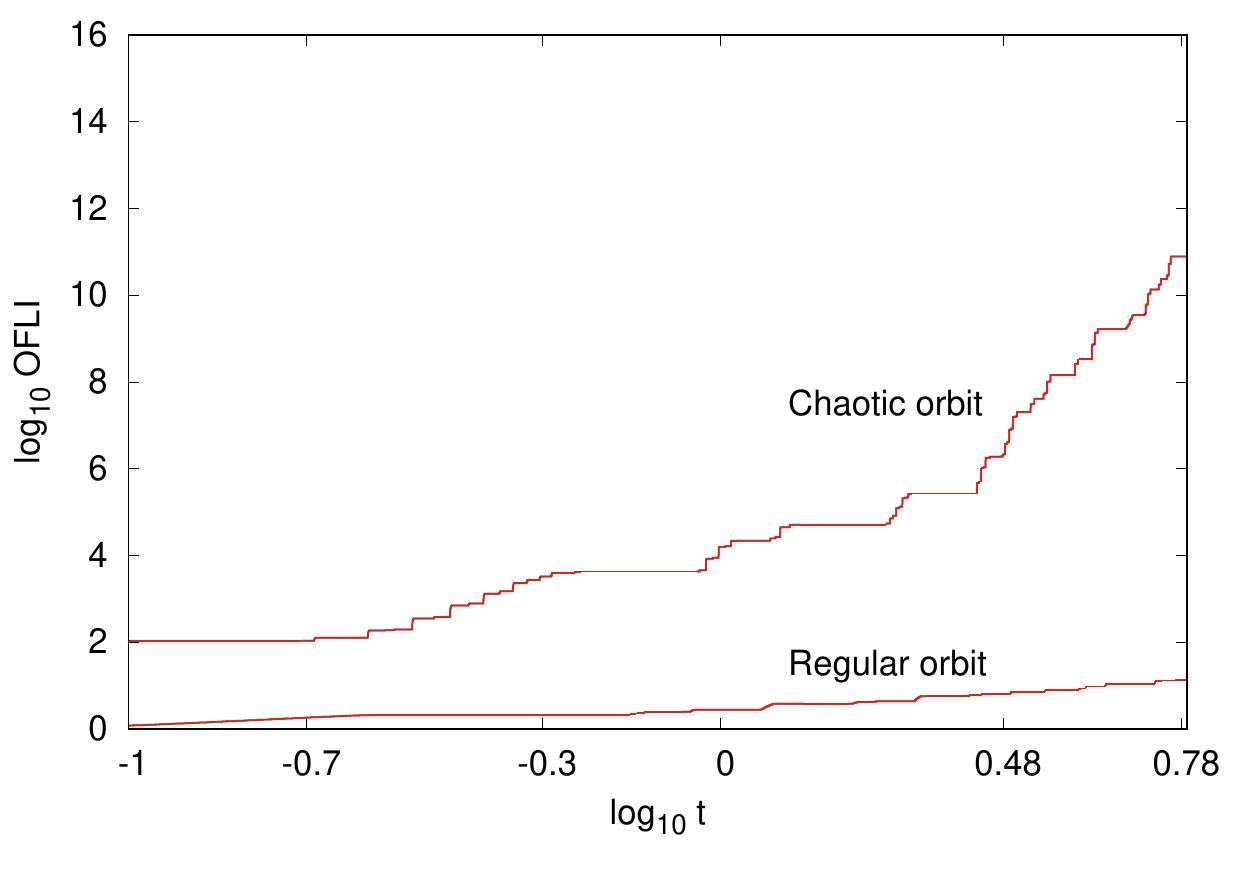}
\caption{Left: Distance $|\Delta \bf{w}|$ in the phase space (in logarithmic scale), as a function of units of time: u.t. (1 u.t. $\sim$ 1 Gyr), between orbits computed with accelerations obtained from {\sc Hydra} 2.0 and with {\sc Smart}. The orbits belong to the time independent flavour of the {\sc Hydra} 2.0 potential, with initial conditions given on Table~\ref{IC} (regular orbit, red and chaotic orbit, blue). Right: The OFLI (a fast chaos indicator) of the regular and the chaotic orbits presented on the left panel. The indicator OFLI should grow linearly or exponentially with time for the regular or the chaotic orbit, respectively. The blue and red curves (almost superposed) correspond to variational equations obtained with {\sc Hydra} 2.0 and {\sc Smart}, respectively. The agreement between both curves and for each orbit is evident.}
\label{TI}
\end{figure}

For both versions of the {\sc Hydra} 2.0 potential, 500 initial conditions sampling the phase-space distribution of the dark matter halo were randomly chosen with which orbits were computed with the {\sc LP-VIcode}, both with the accelerations computed and coded already in {\sc Hydra} 2.0, and with the accelerations obtained with {\sc Smart}. In the case of the time independent version, for all regular examples the orbits thus obtained were the same, aside from the expected numerical differences mentioned in the previous experiments. On the other hand, as previously discussed, chaotic orbits only remain the same until the exponential divergence phase begins to show. The left panel of Fig. \ref{TI} shows the distance in phase space between orbits computed with accelerations coded in {\sc Hydra} 2.0 and those coded by {\sc Smart}, for a regular and a chaotic case (initial conditions are specified on Table~\ref{IC}). We repeated the above procedure for the variational equations coded in {\sc Hydra} 2.0 and coded by {\sc Smart}. With them, we computed all the eleven available chaos indicators with {\sc LP-VIcode} for each orbit. The results were similar to those obtained in the case of accelerations: no significant differences were found between codes for both the regular and chaotic orbits. Furthermore, the regular or chaotic quality was always preserved. For instance, the right panel of Fig. \ref{TI} shows the Orthogonal Fast Lyapunov Indicator, OFLI (a fast chaos indicator, \cite{FLFF02}) applied to the regular and the chaotic case presented in the panel on the left. 

\begin{table}
 \caption{Initial conditions used with  {\sc Hydra} 2.0 for the orbit showed in Fig.~\ref{TI}. Positions are in kpc and velocities in km s$^{-1}$.}
 \label{IC}
 \begin{center}
 \begin{tabular}{crr}
  \hline
  & Regular & Chaotic \\   
  \hline
  $x$ & 0.31807616 & 0.73157591 \\
  $y$ & 6.7984977\phantom{0} & $-0.84585929$  \\
  $z$ & $-20.295328$\phantom{00} & $-0.21206708$ \\
  $\dot x$ & $-92.1442$\phantom{0000} & 8.74071\phantom{000} \\
  $\dot y$ & 244.737\phantom{00000} & 34.4792\phantom{0000} \\
  $\dot z$ & $-86.8133$\phantom{0000} & 119.667\phantom{00000} \\ 
 \hline
 \end{tabular}
 \end{center}
\end{table}

\begin{figure}
\includegraphics[width=0.49\textwidth]{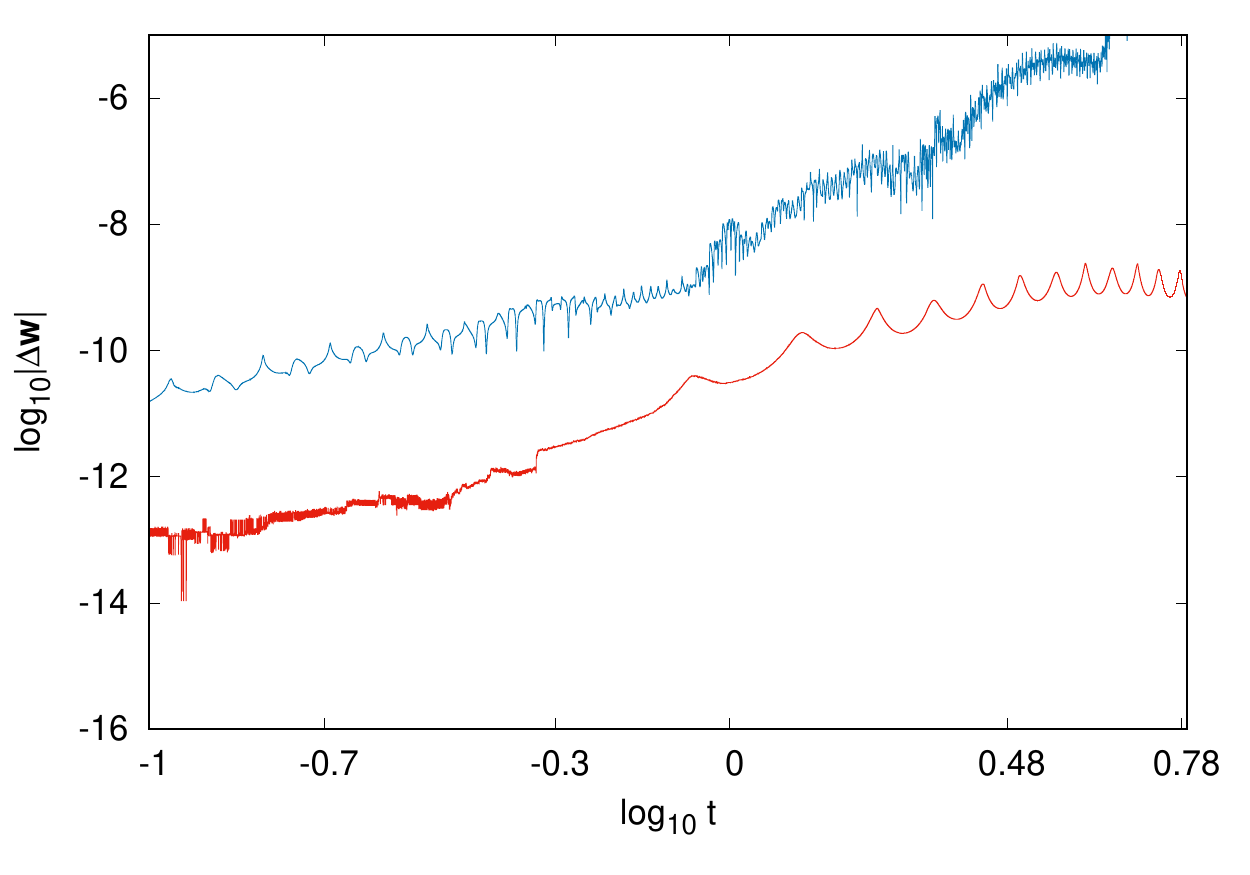}
\includegraphics[width=0.49\textwidth]{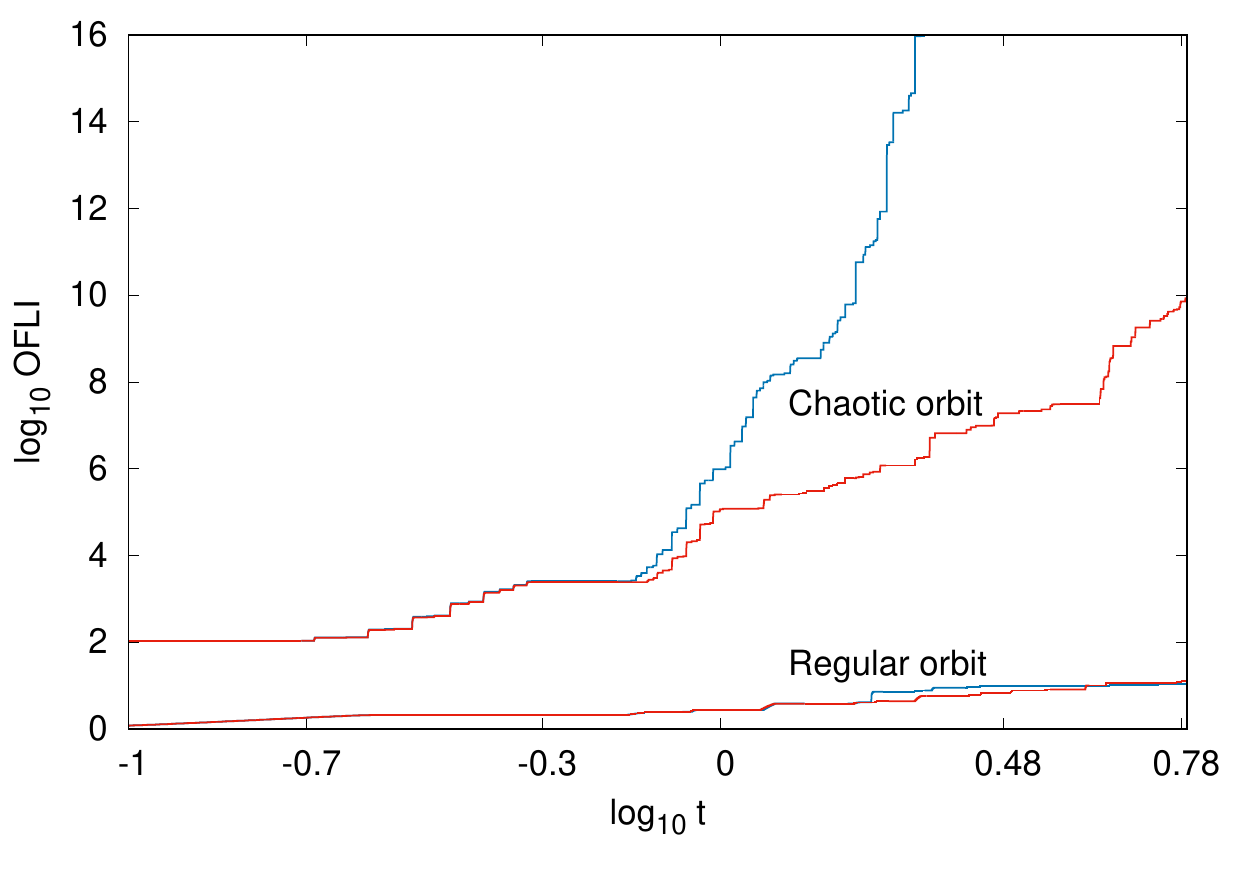}
\caption{Idem Fig.~\ref{TI} but for the full time dependent flavour of the {\sc Hydra} 2.0 potential.}
\label{TD}
\end{figure}

For the full time dependent version of the {\sc Hydra} 2.0, the left panel of Fig. \ref{TD} shows the distance in phase space between orbits computed with accelerations coded in {\sc Hydra} 2.0 and by {\sc Smart}, for the same regular and chaotic orbits presented on Table~\ref{IC}. Once again, for all regular examples the orbits thus obtained were the same, aside from the expected numerical differences, and chaotic orbits remained the same only until the exponential divergence makes the orbits differ. On the right panel of Fig. \ref{TD} we show the OFLI for both regular and chaotic cases, where the accelerations plus the variational equations are those coded in {\sc Hydra} 2.0 (blue) and by {\sc Smart} (red). Both computations for the regular orbit follow very similar trajectories, which is not the case for the chaotic orbit. The separation for both chaotic orbits starts to become noticeable when the time dependent
components of the Milky Way-like potential, i.e. its bar and spiral arms, start to gain strength relative to the background time independent potential. Final amplitudes of the bar and spiral arms are reached by 3.068 units of time, u.t. The results shown for the chaotic orbits are thus expected. In any case, the regular or chaotic quality was preserved also in this experiment.

With this set of experiments we have not only checked that {\sc Smart} is working properly under very demanding circumstances, but also we have checked for a second time and with an independent program ({\sc Smart}) that {\sc MWH} 2.0 has its accelerations and variational equations correctly coded.

We also run for this potential a number of experiments to test the computing times both with {\sc MWH} 2.0 (i.e., the accelerations and variational equations already coded by hand) and {\sc Smart}. In all the experiments {\sc MWH} 2.0 performed faster than {\sc Smart}, an expected result due to the different goals of both codes.

On Table \ref{times} we present the results. The computing times obtained are an average among the four threads running simultaneously for the experiments using big samples of initial conditions. In case of the experiments with only 500 initial conditions, the computing times are the actual times registered for one thread.

\begin{table}
 \caption{Computing times with {\sc MWH} 2.0. The first column indicates whether the version of {\sc MWH} 2.0 is time independent (TI) or the full time dependent version (TD). The second column shows whether the computation of the orbits were done with only the equations of motion (EQM) flag activated; or the EQM and the variational equations (VEQ) with only the MEGNO \& SElLCE and FLI \& OFLI flags activated; or the EQM, the VEQ for all chaos indicators and all other quantities programmed with the {\sc LP-VIcode} (FULL). The third column presents the number of initial conditions integrated in the experiments. The last column shows the time reduction, in percentage, obtained by {\sc MWH} 2.0 compared with {\sc Smart}.}
 \label{times}
 \begin{center}
 \begin{tabular}{llrr}
  \hline
  Flavour & Experiment & Number of & \% \\
  & type & initial conditions&\\
  \hline
  TI & EQM &  1163956 & 7.04 \\
  TD & EQM & 1163955 &  12\phantom{.00} \\
  TD & EQM+VEQ & 180336 &  19\phantom{.00} \\
  TI & FULL & 500 & 31.55 \\
  TD & FULL & 500 & 37.11 \\
 \hline
 \end{tabular}
 \end{center}
\end{table}

It is evident from the experiments that the subprogram {\sc MWH} 2.0 speeds up the computation with respect to {\sc Smart}, due to its more efficient coding. However, such a difference in speed is large only when the computation of variational equations is involved: between 20 per cent and 40 per cent time reduction, depending on the number of chaos indicators computed. 

Therefore, the performance of {\sc Smart} with our Galactic potential is comparable to that obtained with the rest of the potentials. The conclusion is also similar: when chaos indicators on galactic potentials need to be computed on a huge number of orbits, the original version of {\sc MWH} 2.0 is worth using over the automatic differentiation version provided by {\sc Smart}. Otherwise, {\sc Smart} is preferable due to its simplicity.

\subsection{Comparison with other software}

Automatic differentiation can be done in several ways. Perhaps the easiest way to get the expression of a derivative is through modern symbolic manipulators (e.g. \textsc{Mathematica, Maple, Macsyma}, etc.), though if the function to be derived is not simple enough, the resulting expression may be very involved and may require intensive handling to turn it into code form. Numerically, it is always possible to derive a function by using divided differences \citep[e.g.][]{PTVF92}, but this method is inexact and a precise handling of errors becomes necessary. Among the programs that perform the automatic differentiation of a preexisting code, the package \textsc{Taylor} \citep{JZ05} obtains the solution of ordinary differential equations (ODEs) by the Taylor method. By choosing a high enough order in the Taylor series an arbitrary precision can be obtained, but at the cost of increasing the computing time. Also, the ODEs should be written in a specific language to be understood by the program, and it requires a somewhat complex installation procedure. Another software, \textsc{Pcomp} \citep{DLS95} is similar to ours in facing the differentiation of a code, but, as in the case of \textsc{Taylor}, the routine to be derived should be rewritten in a special language. A very general automatic differentiation software, not in the public domain, is \textsc{Adifor 2.0} \citep{BCKM95}, which takes an arbitrary function written in Fortran 77 code and produces its derivative, just as in our case. It overcomes our program in that it can handle simple precision and complex variables, statement functions, derivable variables in \texttt{COMMON} and \texttt{EQUIVALENCE} declarations, and exceptions with non-derivable intrinsic functions. However, in order to include these features a preprocessing step is needed. As already said, most of these features are not necessary in our case, due to the fact that our goal is to derive a double precision continuous potential. Also, the complete package consists of three programs, one of them in \texttt{C}, while our code is a simple program in Fortran. We would have liked to be able to compare the performance of \textsc{Adifor 2.0} with that of our program. Unfortunately, our asking for a license had no answer, so we assume \textsc{Adifor 2.0} is not supported anymore. In contrast to the abovementioned programs, our code does not require any license, does not need installation nor preprocessing, and no new language has to be learned to use it. It is a single F77 program, and all it requires is to compile it and run it, yielding F77 routines ready to use for computing accelerations and variational equations.

\section{Conclusions}

We have developed a program that, given an arbitrary potential written in F77, automatically generates ready-to-use F77 routines to compute its corresponding accelerations (first derivatives) and variational equations (second derivatives). The code of the potential, aside from its heading, may be any valid F77 code except for seven slight limitations, all of which can be bypassed by rewriting the code appropriately. The code was validated with assorted potentials, including time-dependent potentials, velocity-dependent potentials and very complex potentials that even need external routines to be computed. Last but not least, the program has been successfully tested against a realistic, seven-component time-dependent Galactic potential, {\sc MilkyWayHydra}, which produces very involved derivatives and thus is a litmus test for our code.

Funding: DDC acknowledges financial support from the Universidad Nacional de La Plata, Argentina [Proyecto 11/G153]. FAG acknowledges financial support from FONDECYT Regular 1211370 and from the Max Planck Society through a Partner Group grant.

\appendix
\section{The {\sc MilkyWayHydra} potential}
\label{A1}
Current version {\sc MWH} 2.0 describes a multi-component 3D Galactic potential: (i) a central region composed of a nuclear star cluster (NSC), described by a Plummer potential and a supermassive black hole (SMBH) placed in its centre; (ii) a peanut-shaped bulge composed of the bulge spheroid (described by a Hernquist profile) and a bar (described by a three-dimensional quadrupole); (iii) spiral arms, described by a $N$--armed spiral pattern outside of the bar region; (iv) two discs, thin and thick, which can be chosen from two options: a Miyamoto-Nagai (MN) profile or an exponential profile; and (v) a dark matter halo (DMH) that has also two options for choosing it: a bi-triaxial extension of the Navarro--Frenk--White (NFW) model or a modified logarithmic model. 

Early versions of the {\sc MWH} has been successfully used in the following works:  \cite{MDCG13,MGetal15,MGetal18,Setal19,Getal20}. A fully comprehensible document about the {\sc MWH} 2.0 can be found at the website of the {\sc LP-VIsuite} (see Section~\ref{intro}).

\begin{figure}
\includegraphics[width=0.90\textwidth]{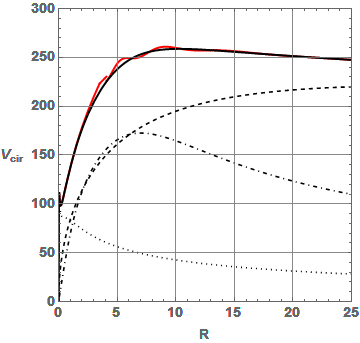}
\caption{Circular velocities of the {\sc Hydra} 2.0 model. Dotted line: nucleus plus spheroidal bulge; dot-dashed line: thin and thick discs; dashed line: dark matter halo; solid black line: full time-independent model (i.e., without bar or arms); solid red line: full model, including bar and arms. Distances expressed in kpc, velocities in km s$^{-1}$.}
\label{velcirc}
\end{figure}

In the following subsections we will describe each component of the {\sc MWH} 2.0 potential. The default values of the parameters found on Table~\ref{param} define the so-called {\sc Hydra} 2.0 potential used for the validation experiment applied to {\sc Smart}, Section~\ref{S4.1}. Fig.~\ref{velcirc} shows the circular velocities generated by the nucleus, the bulge, the discs, and the halo, along with the full curve of the complete model, both with and without the bar and the spiral arms. Table ~\ref{param} also includes the references from which the values of the parameters were taken; those parameters without a reference were chosen to reproduce the circular velocity profile of model Aq-C4/C5 from \cite{MPS14}.

\begin{table}
 \caption{Default values of the parameters defining the {\sc Hydra} 2.0 potential. The parameters of the disc correspond to Option 2 (exponential disc), whereas those of the halo correspond to Option 1 (bi-triaxial NFW profile).}
 \label{param}
 \begin{center}
 \begin{tabular}{llll}
  \hline
  Component & Parameter & Default value & Reference\\
  \hline
  Nuclear region & $M_{\rm nuclear}$ & $2\times10^{8}\,{\rm M}_{\odot}$ & \cite{MGetal18} \\
 & $\epsilon_{\rm nuclear}$ & $0.03\,{\rm kpc}$ & \cite{LZM2002} \\  
Bulge spheroid & $M_{\rm bulge}$ & $4.74\times10^{9}\,{\rm M}_{\odot}$ & \cite{MGetal18}  \\
 & $\epsilon_{\rm bulge}$ & $0.835\,{\rm kpc}$ & \cite{MGetal18} \\
 Bar & $\alpha$ & $0.01$ & \cite{MFSetal16} \\
  & $v_0$ & $220.95\, {\rm km}\,{\rm s}^{-1}$ & \\
 & $R_{\rm b}$ & $3.5\, {\rm kpc}$ & \cite{MFSetal16} \\
  & $R_0$ & $8\, {\rm kpc}$ & \cite{MFSetal16} \\
 & $\Omega_{\rm b}$ & $-52.2\, {\rm km}\,{\rm s}^{-1} {\rm kpc}^{-1}$ & \cite{MFSetal16} \\
 & $\phi_{\rm b}$ & $5.65$ & \cite{MFSetal16}; \\
 &&& \cite{D2000}\\
 Spiral arms & $A$  & $ 2279\, {\rm km}^{2} {\rm s}^{-2}$ & \\
 & $R_{\rm s}$ & $1\, {\rm kpc}$ & \cite{MFSetal16} \\
 & $R_{\rm sd}$ & $3.124\,{\rm kpc}$ & \\
 & $N$ & 2 & \cite{MFSetal16} \\
 & $p$ & $0.17$ &  \cite{MFSetal16}; \\
 &&& \cite{BB2019} \\
 & $\Omega_{\rm s}$ & $-18.9\, {\rm km}\,{\rm s}^{-1} {\rm kpc}^{-1}$ & \cite{MFSetal16}; \\
 &&& \cite{AHDetal2014} \\
 & $\phi_{\rm s}$ & $3.31$ &  \cite{MFSetal16} \\
 & Corrotation radius & $4.08\,{\rm kpc}$ & \\
Thin disc & $ M_{\rm disc}^{\rm thin}$ & $5.276\times10^{10}\,{\rm M}_{\odot}$ & \\
& $R_d^{\rm thin}$ & $3.124\,{\rm kpc}$ & \cite{MGetal18} \\
& $h_z^{\rm thin}$ & $0.3\,{\rm kpc}$ & \cite{MGetal18}\\
Thick disc & $M_{\rm disc}^{\rm thick}$ & $0.686\times10^{10}\,{\rm M}_{\odot}$ & \\
& $R_d^{\rm thick}$ & $3.124\,{\rm kpc}$ & \cite{MGetal18} \\
& $h_z^{\rm thick}$ & $1\,{\rm kpc}$ & \cite{MFSetal16} \\
DMH & $M_{200}$ & $145.64 \times 10^{10}\,{\rm M}_\odot$ & \cite{MGetal18} \\
& $c_{\rm nfw}$ & $16.0287081$ & \cite{MGetal18} \\
& $r_{\rm s}$ & $14.63\,{\rm kpc}$ & \cite{MGetal18} \\
& $(a,b,c)$ & $(1.023,1.010,0.964)$ & \cite{MGetal18}\\
& $(a',b',c')$ & $(1.069,0.984,0.941)$ & \cite{MGetal18} \\
 \hline
 \end{tabular}
 \end{center}
\end{table}

\subsection{The nuclear region}\label{A11}
The nuclear region is described by a Plummer sphere: 

\begin{equation}
\Phi_{\rm nuclear}=-\frac{G\, M_{\rm nuclear}}{\sqrt{x^2+y^2+z^2+\epsilon^2_{\rm nuclear}}},
\label{nuclear}
\end{equation}
where $M_{\rm nuclear}$ is the total mass of the nuclear region, which includes the NSC and the mass of the SMBH, and $\epsilon_{\rm nuclear}$ is the scale length of the system, which defines the outer limit of the inner core.

\subsection{The peanut-shaped bulge}\label{A12}
The abovementioned nuclear region dominates the innermost parts of the Galaxy. The stellar bulge, on the other hand, dominates the region between 0.4 and 3 kpc, approximately \cite{LZM2002}. The bulge spheroid is described by a Hernquist profile:

\begin{equation}
\Phi_{\rm bulge}=-\frac{G\, M_{\rm bulge}}{\sqrt{x^2+y^2+z^2}+\epsilon_{\rm bulge}},
\label{bulge}
\end{equation}
where $M_{\rm bulge}$ is the mass, and $\epsilon_{\rm bulge}$ is the scale length.

The bar potential is a 3D version taken from \cite{MFSetal16} of the pure quadrupole model used by, e.g. \cite{D2000}:

\begin{equation}
\Phi_{\rm bar}(R,\phi,z,t)=\alpha\frac{ v_0^2}{3}\left(\frac{R_0}{R_{\rm b}}\right)^3U(r)\frac{R^2}{r^2}\cos(\gamma_{\rm b}),
\label{bar}
\end{equation}
where $R^2 = x^2 + y^2$ is the radial coordinate on the Galactic plane, $r^2 = R^2 + z^2$ is the spherical radius, $R_{\rm b}$ is the length of the bar, $R_0$ is the Galactocentric radius of the Sun, and $v_0$ is the circular velocity at $R_0$. The amplitude $\alpha$ is the ratio between the bar's and the axisymmetric contribution to the radial force along the bar's long axis at $(R, z) = (R_0, 0)$. The parameter $\gamma_{\rm b}(\phi,t)= 2\left(\phi-\phi_{\rm b}-\Omega_{\rm b}\;t\right)$, where $\phi_{\rm b}$ and $\Omega_{\rm b}$ are the initial phase and angular velocity of the bar pattern, respectively. For simulations that include the present time, it should be taken into account that $\phi_{\rm b,present}=-25^\circ$. Finally,  $U(r)$ is given by:

\begin{equation}
U(r)=\left\{ \begin{array}{lcl}
-(r/R_{\rm b})^{-3} & {\rm for } & r\ge R_{\rm b},\\
(r/R_{\rm b})^{3}-2 & {\rm for } & r< R_{\rm b}.
\end{array}
\right.
\label{bar2}
\end{equation}

\subsection{The spiral arms}\label{A13}
Following \cite{MFSetal16}, we describe the non-axisymmetric part of the Milky Way disc by a $N$-armed spiral pattern outside of the bar region, in the form originally proposed by \cite{CG2002}:

\begin{equation}
\Phi_{\rm spiral}(R,\phi,z,t)=-\frac{A}{R_{\rm s} K D}\cdot {\rm e}^{-\frac{\left(R-R_{\rm s}\right)}{R_{\rm sd}}}\cos(\gamma_{\rm s})\left[{\rm sech}\left(\frac{K z}{\beta}\right)\right]^\beta,
\label{spiral}
\end{equation}
where
\begin{equation}
\begin{split}
K(R)&=\frac{N}{R\sin p},\\
\beta(R)&=K(R) h_{\rm s}\left[1+0.4K(R) h_{\rm s}\right],\\
D(R)&=\frac{1+K(R) h_{\rm s}+0.3\left[K(R) h_{\rm s}\right]^2}{1+0.3 K(R) h_{\rm s}},\\
\gamma_{\rm s}(R,\phi,t)&=N\left[\phi-\phi_{\rm s}-\Omega_{\rm s} t-\frac{\ln(R/R_{\rm s})}{\tan p}\right].
\end{split}
\label{spiral2}
\end{equation}
In Eqs. (\ref{spiral}) and (\ref{spiral2}), $p$ is the pitch angle, $A$ the amplitude of the spiral potential, $h_{\rm s}$ controls the scale-height of the spiral, $R_{\rm s}$ is the reference radius for the angle of the spirals, $R_{\rm sd}$ is a length parameter that must be in agreement with the scale-length of the disc, $N=2$ in order to have a two--armed spiral pattern (namely, the Scutum-Centaurus and Perseus arms), and $\phi_{\rm s}$ and $\Omega_{\rm s}$ are the initial phase and angular velocity of the spiral pattern, respectively (once again, those values are computed using a final integration time $t_e$). As in the case of the bar, the initial phase should be computed so that, at the present, the arms have their actual position.

The spiral arms representation is discarded inside the corrotation radius due to two main reasons: not only the model is poorly adequate for those inner locations (because of the multipolar expansion), but also it is a second-order perturbation that would go inside a region dominated by the bar. 

A method to generalize the spiral pattern representation given by Eqs. (\ref{spiral}) and (\ref{spiral2}) is explained in \cite{CG2002}. The user can add more modes by playing around with the values of $n$ (number of modes) and $N$ (number of arms) in the expression:

\begin{equation}
\Phi_{\rm spiral}(R,\phi,z,t)=-\frac{A}{R_{\rm s}}\cdot {\rm e}^{-\frac{\left(R-R_{\rm s}\right)}{R_{\rm sd}}}\sum_{i=1}^n\frac{C_i}{K_i D_i}\cos(\gamma_{{\rm s},i})\left[{\rm sech}\left(\frac{K_i z}{\beta_i}\right)\right]^{\beta_i},
\label{spiral3}
\end{equation}
where
\begin{equation}
\begin{split}
K_i(R)&=\frac{i N}{R\sin p},\\
\beta_i(R) &=K_i(R) h_{\rm s}\left[1+0.4 K_i(R) h_{\rm s}\right],\\
D_i(R)&=\frac{1+K_i(R) h_{\rm s}+0.3\left[K_i(R)h_{\rm s}\right]^2}{1+0.3 K_i(R)h_{\rm s}}\\
\gamma_{{\rm s},i}(R,\phi,t)&= i N\left[\phi-\phi_{\rm s}-\Omega_{\rm s} t-\frac{\ln(R/R_{\rm s})}{\tan  p}\right],
\end{split}
\label{spiral4}
\end{equation}
and $C_i$ are coefficients that set the power of each mode. Thus, the series is a sum of terms, each one being similar to the two-armed potential of Eqs. (\ref{spiral}) and (\ref{spiral2}). Then, since all the first and second derivatives are the same for all those terms (except for the constants), the user can build a general $N$-armed potential with its equations of motion and first variational equations by simply (but carefully) adding those expressions.

\subsection{The disc}\label{A14}
The {\sc MWH} 2.0 offers two different options to represent the disc:

\begin{enumerate}
\item Option 1: a MN profile offers analytical and fairly simple expressions for the first and second derivatives of the disc potential in order to compute the equations of motion as well as the first variational equations;  

\item Option 2: an exponential profile for the mass distribution had to be avoided, given that the corresponding potential includes Bessel functions that need to be integrated numerically. Instead, we have used a combination of MN models to build disc potentials that approximates such exponential mass profiles (following \cite{SFC2015}) and keeping at the same time the analytical and rather simple form of the first and second derivatives. 
\end{enumerate}

We first introduce the well-known MN disc potential that is used for both options coded in {\sc MWH} 2.0 to build up the disc profiles:
 
\begin{equation}
\Phi_{\rm disc}=-\frac{G\,M_{\rm disc}}{\sqrt{x^2+y^2+\left (\epsilon_{s}+\sqrt{z^2+\epsilon^2_{h}}\right)^2}},
\label{disc}
\end{equation}
where $M_{\rm disc}$ is its mass, and $\epsilon_{s}$ and $\epsilon_{h}$ are the scale length and scale height of the disc, respectively. 

Option 1 can be turned into a two-component MN disc in a very easy way: to the default MN disc that is already coded in the subprogram {\tt milkywayhydra.pav}, the user may simply add the blocks corresponding to the second disc by copying and pasting those of the first.

On the other hand, Option 2 uses a combination of MN discs to build a double exponential mass profile for each single disc. The density profile is given by: 

\begin{equation}
\rho(R,z)=\rho_0\exp(-R/R_{\rm d})\exp(-|z|/h_z),
\label{disc2}
\end{equation}
where the double-exponential is described by the parameters $\rho_0$, $R_{\rm d}$, and $h_z$, which are the central density, the scale length and the scale height, respectively. Thin and thick exponential discs are already coded in the subprogram {\tt milkywayhydra.pav}, following the work of \cite{SFC2015}. The procedure is explained in detail in the {\sc MWH} 2.0 descriptive memory that is freely available in the website of the suite (Section~\ref{intro}).

\subsection{The dark matter halo}\label{A16}
Finally, we introduce the hypothetical outermost galactic component: the DMH. The {\sc MWH} 2.0 offers two different options to represent such component:

\begin{enumerate}
\item Option 1: a bi-triaxial extension of the NFW DMH used in \cite{VWHS08}. They introduce a transition scale to have a triaxial profile (defined by an ellipsoidal radius) in the inner regions while leaving the outer regions round shaped as dark matter-only simulations have showed. In \cite{MGetal18} we applied a second parameter to introduce triaxiality in the outer parts as well (this giving the bi-triaxial nature to the component). The idea is to have an oblate profile in the inner regions due to the impact of baryonic matter, while leaving the outer regions mildly triaxial as hydrodynamic simulations show, \cite{MPS14,Getal17};

\item Option 2: a modified logarithmic DMH formerly used in \cite{VH2013}. The authors fitted the observations of the Sagittarius dwarf galaxy \cite{LM2010} guaranteeing at the same time a stable configuration for the disc. To this end, they also used a transition scale to obtain an oblate shape in the inner parts (as hydrodynamic simulations suggest, \cite{MPS14}) and a triaxial shape in the outer parts because of the effect of the Sagittarius dwarf galaxy orbit.
\end{enumerate}

Option 1 is described by the following expression:

\begin{equation}
\Phi_{\rm DMH}=-\frac{A_d}{r^\prime}\ln\left[1+\frac{r^{\prime}}{r_{\rm s}}\right],
\label{dmh}
\end{equation}
where
\begin{equation}
A_d=\frac{G\,M_{200}}{\ln(1+c_{{\rm nfw}})-\frac{\displaystyle{c_{\rm nfw}}}{\displaystyle{1+c_{\rm nfw}}}}.
\end{equation}
Here $M_{200}$ is the halo virial mass (defined to be the mass inside a sphere in which the mean matter density is 200 times the critical density, $\rho_{cri}=3H^2(z)/(8\pi G)$ with $H$ the Hubble parameter), $c_{\rm nfw}$ is the concentration parameter, and $r^\prime= r_{\rm ie}\cdot (r_{\rm s}+r_{\rm oe})/(r_{\rm s}+r_{\rm ie})$ 
is the modified scale radius adapted from \cite{VWHS08}, with $r_{\rm s}=r_{200}/c_{\rm nfw}$ a scale radius and $r_{200}$ the virial radius. Also, $r_{\rm ie}^2=x^2/a^2+y^2/b^2+z^2/c^2$ is the square of the inner ellipsoidal radius with $a$, $b$, and $c$ the inner semi-axes, while $r_{\rm oe}^2=x^2/a'^2+y^2/b'^2+z^2/c'^2$ is its outer counterpart, with $a'$, $b'$, and $c'$ the outer semi-axes. The ellipsoidal radii satisfy $a^2+b^2+c^2=a'^2+b'^2+c'^2=3$.

Option 2: this form for the DMH potential describes 
a triaxial ellipsoid with its intermediate and major axes rotated with respect to the coordinate axes about the galactic $z$ axis. The expression of the potential is:
 
\begin{equation}
\Phi_{\rm dmh}=v_{\rm halo}^2\ln[(r^{\star})^2+d^2].
\label{dmh2}
\end{equation}
Here $v_{\rm halo}$ is the halo mass renormalization parameter that should be calibrated in order to obtain the circular velocity of the Sun $v_{\rm LSR}(R_0)$. The radius $r^\star$ is defined as  $r^\star=r_{\rm ie}\cdot (r_{\rm s}+r_{\rm oe})/(r_{\rm s}+r_{\rm ie})$, where now $r_{\rm ie}^2=x^2+y^2+z^2/q_z^2$ and $r_{\rm oe}^2=W x^2+V y^2+U xy+z^2/q_3^2$. Here the parameters $q_z$ and $q_3$ are the axial flattenings perpendicular to the galactic disc, and $W =a_1^2/q_1^2+a_2^2/q_2^2$, $V =a_1^2/q_2^2+a_2^2/q_1^2$, $U = 2 a_1 a_2 (1/q_1^2-1/q_2^2)$, where $q_1$ and $q_2$ are the axial flattenings along the equatorial axes, $a_1=\cos\phi$, $a_2=\sin\phi$, and $\phi$ is the angle of rotation of the potential around the $z$ axis. Finally, $d$ is the core radius of the logarithmic potential. 

Before leaving the subject of the {\sc MWH} components, a comment on the stellar halo is worth mentioning. The stellar halo may be oblate, strengthening the chaotic effects given by the triaxiality of the DMH \cite{DBE2011}. However, its mass is about 1 per cent of the stellar component of the galaxy, and thus for all except the most specialized studies it can be neglected. This is the main reason why the current version of {\sc MWH} does not include such a stellar halo component.

\bibliographystyle{model2-names}
\bibliography{biblio.bib}







\end{document}